\documentclass[twocolumn,aps,prl,superscriptaddress]{revtex4}

\usepackage[dvips]{graphicx}

\begin{document}
\voffset=0.5 in

\title{Pion-Kaon correlations in central Au+Au collisions 
at $\sqrt{s_{NN}} = 130$ GeV}

%

\affiliation{Argonne National Laboratory, Argonne, Illinois 60439}
\affiliation{Brookhaven National Laboratory, Upton, New York 11973}
\affiliation{University of Birmingham, Birmingham, United Kingdom}
\affiliation{University of California, Berkeley, California 94720}
\affiliation{University of California, Davis, California 95616}
\affiliation{University of California, Los Angeles, California 90095}
\affiliation{Carnegie Mellon University, Pittsburgh, Pennsylvania 15213}
\affiliation{Creighton University, Omaha, Nebraska 68178}
\affiliation{Nuclear Physics Institute AS CR, \v{R}e\v{z}/Prague, Czech Republic}
\affiliation{Laboratory for High Energy (JINR), Dubna, Russia}
\affiliation{Particle Physics Laboratory (JINR), Dubna, Russia}
\affiliation{University of Frankfurt, Frankfurt, Germany}
\affiliation{Indiana University, Bloomington, Indiana 47408}
\affiliation{Insitute  of Physics, Bhubaneswar 751005, India}
\affiliation{Institut de Recherches Subatomiques, Strasbourg, France}
\affiliation{University of Jammu, Jammu 180001, India}
\affiliation{Kent State University, Kent, Ohio 44242}
\affiliation{Lawrence Berkeley National Laboratory, Berkeley, California 94720}\affiliation{Max-Planck-Institut f\"ur Physik, Munich, Germany}
\affiliation{Michigan State University, East Lansing, Michigan 48824}
\affiliation{Moscow Engineering Physics Institute, Moscow Russia}
\affiliation{City College of New York, New York City, New York 10031}
\affiliation{NIKHEF, Amsterdam, The Netherlands}
\affiliation{Ohio State University, Columbus, Ohio 43210}
\affiliation{Panjab University, Chandigarh 160014, India}
\affiliation{Pennsylvania State University, University Park, Pennsylvania 16802}
\affiliation{Institute of High Energy Physics, Protvino, Russia}
\affiliation{Purdue University, West Lafayette, Indiana 47907}
\affiliation{University of Rajasthan, Jaipur 302004, India}
\affiliation{Rice University, Houston, Texas 77251}
\affiliation{Universidade de Sao Paulo, Sao Paulo, Brazil}
\affiliation{University of Science \& Technology of China, Anhui 230027, China}
\affiliation{Shanghai Institute of Nuclear Research, Shanghai 201800, P.R. China}
\affiliation{SUBATECH, Nantes, France}
\affiliation{Texas A \& M, College Station, Texas 77843}
\affiliation{University of Texas, Austin, Texas 78712}
\affiliation{Valparaiso University, Valparaiso, Indiana 46383}
\affiliation{Variable Energy Cyclotron Centre, Kolkata 700064, India}
\affiliation{Warsaw University of Technology, Warsaw, Poland}
\affiliation{University of Washington, Seattle, Washington 98195}
\affiliation{Wayne State University, Detroit, Michigan 48201}
\affiliation{Institute of Particle Physics, CCNU (HZNU), Wuhan, 430079 China}
\affiliation{Yale University, New Haven, Connecticut 06520}
\affiliation{University of Zagreb, Zagreb, HR-10002, Croatia}
\author{J.~Adams}\affiliation{University of Birmingham, Birmingham, United Kingdom}
\author{C.~Adler}\affiliation{University of Frankfurt, Frankfurt, Germany}
\author{M.M.~Aggarwal}\affiliation{Panjab University, Chandigarh 160014, India}
\author{Z.~Ahammed}\affiliation{Purdue University, West Lafayette, Indiana 47907}
\author{J.~Amonett}\affiliation{Kent State University, Kent, Ohio 44242}
\author{B.D.~Anderson}\affiliation{Kent State University, Kent, Ohio 44242}
\author{M.~Anderson}\affiliation{University of California, Davis, California 95616}
\author{D.~Arkhipkin}\affiliation{Particle Physics Laboratory (JINR), Dubna, Russia}
\author{G.S.~Averichev}\affiliation{Laboratory for High Energy (JINR), Dubna, Russia}
\author{S.K.~Badyal}\affiliation{University of Jammu, Jammu 180001, India}
% ...

\collaboration{STAR Collaboration}\homepage{www.star.bnl.gov}\noaffiliation

\begin{abstract}

 Pion-kaon correlation functions are constructed from  central Au+Au STAR data taken at
$\sqrt{s_{NN}} = 130$ GeV by the STAR detector at the Relativistic Heavy Ion Collider (RHIC).  
The results suggest that pions 
and kaons are not emitted at the same average space-time point. 
Space-momentum correlations, i.e. 
transverse flow, lead to a space-time emission asymmetry of pions and kaons
that is consistent with the data.  This result provides new independent
evidence that the system created at RHIC undergoes a 
collective transverse expansion.

\end{abstract}


\maketitle


Two-particle correlations for non-identical particles produced in heavy ion collisions are sensitive 
to differences in the average emission time and position of the different particle species~\cite{PLBNonId}. 
Such correlations in data taken at GANIL ($^{129}$Xe+$^48$Ti at 45 MeV per nucleon) 
suggest delayed emission of deuterons with respect to 
protons~\cite{LedNonId}. Correlation data from the SPS (Pb-Pb collisions at $\sqrt{s_{NN}} = 17.3$ GeV), 
and AGS (Au-Au collisions at $\sqrt{s_{NN}} = 4.7$ GeV) also suggest that the pion and 
proton average space-time emission points do not coincide~\cite{LedNonId,NA49,AGS}; a partial explanation 
is that space-momentum correlations arise from the system's collective expansion~\cite{LedNonId}. For Au+Au 
collisions at $\sqrt{s_{NN}} = 130$ GeV, transverse mass 
spectra, elliptic flow, and deduced pion source radii suggest
collective expansion in the transverse plane~\cite{BlastWave, PidFlow}.
Such transverse flow 
may shift the average emission radii of different particle species by different amounts. 
Also, different species may kinematically decouple from the system at different times depending upon their interaction cross sections~\cite{NuSorge}. In addition, the average emission time for a given
species may be delayed significantly if produced dominantly through 
resonance decay. We construct pion-kaon correlation functions from Au+Au STAR data taken
 at $\sqrt{s_{NN}} = 130$ GeV and investigate whether the pions and kaons are emitted at the same average 
space-time position.

Non-identical charged particles interact through Coulomb and strong interactions;
for the pion-kaon case correlation effects are dominated by the Coulomb interaction. 
To probe the $\pi$-$K$ separation, 
correlation functions $C(k^*)$ are constructed as the ratio of the
$k^*$ distribution constructed with particles from the same event (correlated
distribution) divided by the
$k^*$ distribution constructed with particles from different events (uncorrelated
distribution). 
$k^*$  is the magnitude of the three-momentum of either particle 
in the pair rest frame. 

For two particles initially moving towards each other the effects of the Coulomb and strong 
interactions are different from those for two particles initially moving apart.
The technique exploits
this difference to study emission asymmetries~\cite{PLBNonId,PRLNonId,LedNonId} . 
Pairs are divided into two groups, which represents either the case
where the pions catch up with the kaons or the case where the pions move away
from the kaons, depending upon the space-time separation between pion and
kaon emission points.
Each sample is used to construct
two different correlation functions, $C_+(k^*)$ and $C_-(k^*)$, the sign index reflecting
the sign of $\overrightarrow{v}$$\cdot$$\overrightarrow{k^*_{\pi}}$, with 
$\overrightarrow{v}$  the pair velocity and  $\overrightarrow{k^*_{\pi}}$
the pion momentum vector in the pair rest frame. 
If the average space-time emission points of pions and kaons
coincide, both correlation functions are identical. If instead, 
pions are emitted closer to the center of the source than kaons, pions
with larger velocity will tend to catch up with kaons, 
and the Coulomb correlation strength will be enhanced compared to the  case where pions are slower
than kaons.
Hence, the correlation function $C_+$ will show a larger deviation from
unity than $C_-$. 
Pairs can be 
separated according to the sign of 
$k^*_{side}$, $k^*_{long}$ and $k^*_{out}$,
the $\overrightarrow{k^*_{\pi}}$ projections
onto three perpendicular axes in the longitudinally
 comoving system (LCMS) where the longitudinal component
of the pair momentum vanishes~\cite{NonIdOutSideLong}. The $out$ axis parallels
the pair velocity in the LCMS, the $long$ axis is the beam axis
and the $side$ axis is perpendicular to the other two.
 $r^*_{out}$,  $r^*_{side}$, and  $r^*_{long}$ are the corresponding projections of the three-vector 
$\overrightarrow{r}^*$, the relative distance between the particle freeze-out  points 
in the pair rest frame. 
Due to azimuthal symmetry and symmetry about mid-rapidity,
$\langle r^*_{side} \rangle= \langle r^*_{long} \rangle = 0$. Thus 
$C_+/C_-$ defined with respect to the signs of  $k^*_{side}$ and $k^*_{long}$
must equal  unity.
If pions and kaons are
not emitted at the same average radius in the transverse plane and/or 
at the same average time, $C_+/C_-$  defined with respect to the sign of
$k^*_{out}$ will deviate from unity, unless these two contributions cancel. Thus,
one can probe the space-time separation between 
pion and kaon sources in the transverse plane.


Charged particles are identified and tracked by the 
STAR Time Projection Chamber (TPC)~\cite{STARTpc}.
This analysis selects the $12\%$ most central collisions, i.e. the
events with the largest multiplicity of particles. 
Selected particles have pseudorapidity $|\eta| < 0.5$.
The Au+Au collision point (primary vertex) is required to be
within $\pm 75$ cm of the TPC mid-plane.
The non-correlated pair background
 is constructed by mixing events whose primary vertices are also 
separated from each other by less than 10 cm.

Pions and kaons are identified by measuring  specific energy loss ($dE/dx$) in the 
TPC. When the momentum of pions
and kaons exceeds 700 MeV/c, the $dE/dx$ of both species becomes similar
which compromises particle identification. In addition, the
 pion and kaon samples are contaminated by electrons and
positrons. 
The yield of each particle species in the momentum 
range where the energy losses coincide is interpolated  ($e^+/e^-$ 
contamination) or extrapolated (kaon/pion separation) 
from the yields measured in the momentum range where there is good separation.
 In order to quantify the probability of correctly identifying a given
species when the $dE/dx$ bands overlap, four probabilities are calculated
for each track: 
 the chance that the particle is a  $\pi^+$ or $\pi^-$, $K^+$ or $K^-$, 
$p$ or $\overline{p}$, or $e^+$ or $e^-$~\cite{PidFlow}.  To be accepted as a pion or kaon the probability has to be $>$~60\%.
Tracks must point back to within 3 cm of the primary vertex; this
removes a large number of secondary pions.
Pions must have transverse momentum $80$ MeV/c $< p_T < 250$ MeV/c and rapidity $|y|<0.5$, while kaons must have 
 $400$ MeV/c $< p_T < 700$ MeV/c, and $|y|<0.5$.  

Pion-kaon pair identification probability (product of both particle individual
$dE/dx$ probabilities) is required to be larger than 60\%. 
Since the $e^+$-$e^-$ pairs can distort
$\pi^-$-$K^+$ and $\pi^+$-$K^-$ correlation functions, the 
maximum probability allowed for a given pair to be $e^+$-$e^-$  is
set at 1\%, ensuring negligible contribution.  
Track pairs that share 
more than 10\% of their TPC space points are discarded 
in order to avoid  track merging errors. 
Two points are defined as shared if
the probability of separating 
hits produced by them in the TPC is less than 99\%.
After selecting pion-kaon pairs, the correlation functions are constructed by taking the  ratio of the $k^*$ distributions of pairs from the same event to the 
$k^*$ distributions of pairs from different events.


Primary purity and momentum resolution effects are taken into account as described below. 
Primary purity is the percentage of primary pion-kaon pairs in all
pion-kaon pairs satisfying all cuts. It is
estimated to average $77\%$ for unlike sign pairs and 
$75\%$ for like sign pairs. The lower limit for each is $54\%$.  This number is the product of 
the probability of identifying both pions and kaons
using the $dE/dx$ information and the probability of  excluding pions and kaons that do not
originate from points close to the collision vertex.  Excluded
pions include
decay products of strange hyperons and $K^0_s$, and pions produced
in the detector material. 
The fraction of secondary pions is estimated from the $K^0_s$,
$\Lambda$ and pion yields in ~\cite{PhenixSpectra, K0,La}. 
Detector simulations
determine the relative reconstruction efficiency of pions from these different sources.
Secondary kaons, being rare,  are neglected.
Assuming that the non-primary pion-kaon pairs are
uncorrelated, the correlation function is corrected as
$C_{true}(k^*) = (C_{measured}(k^*) - 1)/purity(k^*)+1$. The systematic error
introduced by this correction is  less than $20\%$.


The effect of momentum resolution depends upon the correlation function shape.
Pion-kaon correlation functions are calculated from the pion and kaon momentum and space-time 
distributions, accounting
for both the Coulomb and strong interactions as in 
 ~\cite{LedLubo}. The correlation function strength 
is calculated with the true momentum while the correlation 
function is binned as a function of $k^*$ smeared by momentum resolution.
Momentum resolution is estimated at the track level by detector simulations.
The space-time distribution is chosen so that the main features 
of the measured correlation function are reproduced.
The correction
is obtained by comparing  correlation functions calculated with and without momentum smearing. 
The correction enhances $C(k^*)$ by 20\% 
(1\%) for
$k^* < 5$ MeV/c ($5 < k^* < 10$ MeV/c), first and second bins in Figure 1, with a conservative systematic
error of $\pm100\%$ on the correction of these two bins.


\begin{figure}[ht]
\includegraphics[width=.48\textwidth]{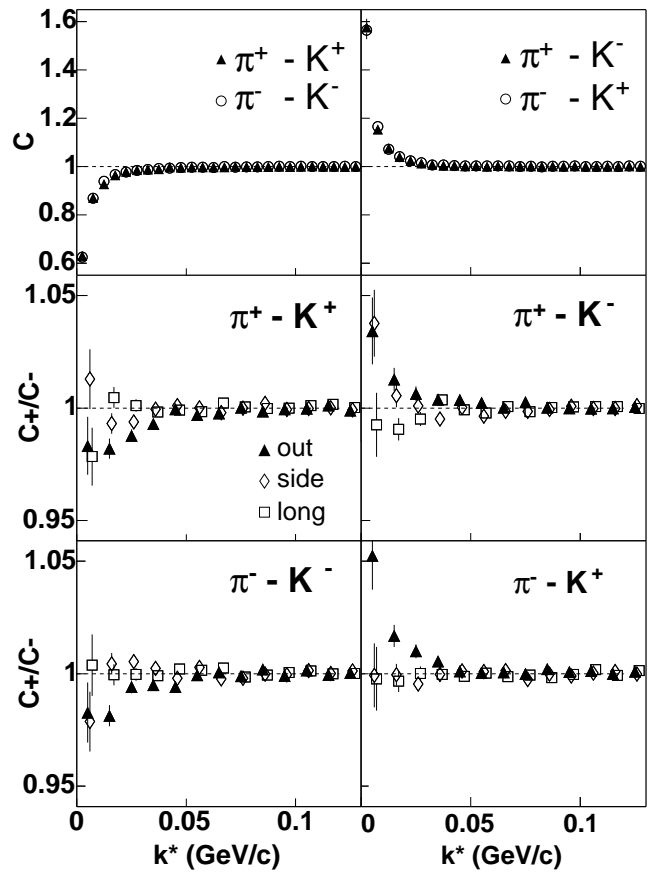}
\caption{\label{Fig1}  Top panels: pion-kaon correlation functions
$C(k^*)$, the average of $C_+(k^*)$ and $C_-(k^*)$.
Middle and bottom  panels: ratio of the correlation functions $C_+$ and $C_-$ 
defined with the sign of the projections, $k^*_{out}$, 
$k^*_{side}$ and $k^*_{long}$. 
Errors are statistical only. The horizontal axis of the ratios $C_+/C_-$ 
for $k^*_{side}$  ($k^*_{long}$) is shifted by  1 MeV/c (2MeV/c) to separate 
the error bars.}
\end{figure}

The top panels of Figure 1 show the correlation functions for every 
combination of 
pion-kaon pairs.  
The agreement between unlike-sign ($\pi^-$-$K^+$ and $\pi^+$-$K^-$)
and between like-sign ($\pi^+$-$K^+$ and $\pi^-$-$K^-$) correlation
functions is excellent. 
The middle and bottom panels show the ratios $C_+/C_-$ 
for all pion-kaon pair 
combinations.  $C_+/C_-$ with respect to 
the sign of  $k^*_{side}$ and $k^*_{long}$  is unity within statistical 
errors in accordance with the requirement that $\langle r^*_{side} \rangle = 
\langle r^*_{long} \rangle = 0$.
However, $C_+/C_-$ with respect to the sign of  $k^*_{out}$
is significantly larger than unity at small $k^*$ when the interaction is attractive ($\pi^-$-$K^+$ and $\pi^+$-$K^-$) and significantly  smaller than unity when the  interaction is repulsive  ($\pi^+$-$K^+$ and $\pi^-$-$K^-$). 
These results indicate that pions and kaons are not emitted on average at the same radius and/or time. 


\begin{figure}[ht]
\includegraphics[width=.48\textwidth]{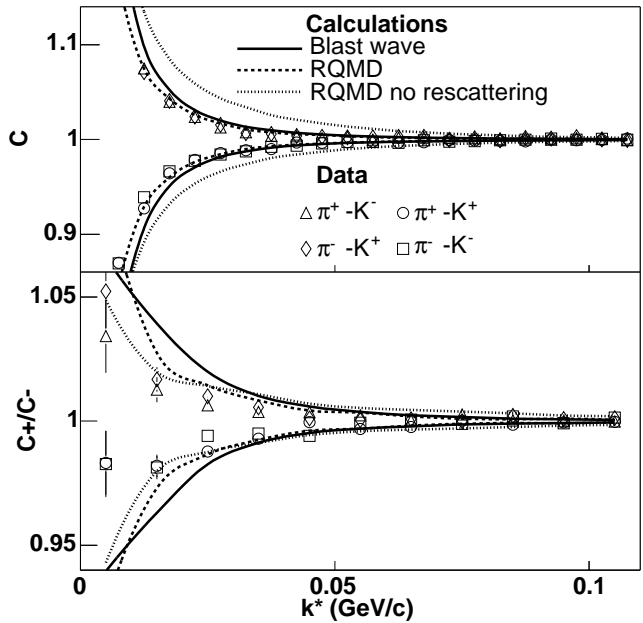}
\caption{\label{Fig2} Comparison of the correlation functions
between data and model. Upper panel, $C(k*)$ correlation function.
Lower panel, ratio  $C_+/C_-$ with respect to 
the sign of  $k^*_{out}$.}
\end{figure}

In order to understand the measured average space-time shift between
pion and kaon sources, we
compare the data with the RQMD (Relativistic Quantum Molecular Dynamic~\cite{RQMD}) model and
the Blast Wave Parametrization (BWP) described in Ref.~\cite{BlastWave}. 
BWP assumes that the system has 
undergone longitudinal and transverse expansions, and provides
the particle space-time and momentum distributions at kinetic freeze-out. 
The parameters, system outermost radius $R = 13$ fm, 
freeze-out proper  time $\tau = 9$ fm/c, emission duration $\Delta\tau = 0$ fm/c, temperature $T=110$ MeV,  and
transverse flow rapidity $\rho(r) =  0.9 (r/R)$ (with  particle emission radius $r$)  
are consistent with fits to pion, 
kaon, proton and lambda transverse mass spectra and to pion source radii~\cite{BlastWave}. 
The hadronic cascade model, RQMD, also generates transverse flow through 
rescattering of hadrons~\cite{NuSorge}. Indeed, turning off hadronic rescattering within this model shuts off transverse flow~\cite{KaisPaper}.
In addition, RQMD includes
contributions from resonance decay, such as $\omega$, $\eta$ and $\phi$, which
shift pion and kaon emission times.  


Figure 2 shows 
correlation functions $C(k^*)$ and ratios $C_+/C_-$ 
measured, from
BWP, and from RQMD with and without hadronic rescattering. 
The calculated correlation functions use model space-time and momentum distributions as described in
~\cite{LedLubo},  with pion
    and kaon kinematic cuts chosen to match the data. 
The correlation function are calculated for like-sign and unlike-sign pairs  
The small wiggles in the calculated $C_+/C_-$ ratios for $k^* < $ 20 MeV/c arrise from statistical uncertainties.
RQMD and BWP are in qualitative agreement
with the measured correlation functions. Turning off rescattering 
in RQMD leads to a  strong correlation, which implies that the pion
and kaon sources are too small. On the other hand, 
RQMD reproduces qualitatively the ratio $C_+/C_-$.



The effect of source size and source shift is disentangled by simultaneously fitting the correlation functions
$C_+$ and $C_-$.
In order to insure that the detector acceptance is matched, the particle momenta are taken from experimental pion-kaon pairs constructed by mixing events
that pass all the cuts.  
The particle positions are set such that the distribution of the relative space-time separation between pions and kaons in the pair rest frame is a three dimensional Gaussian. The free parameters are the Gaussian mean, $\langle \Delta r^*_{out} \rangle = 
\langle r^*_{out}(\pi) - r^*_{out}(K) \rangle$ ($\langle \Delta r^*_{side} \rangle =  \langle \Delta r^*_{long} \rangle =0$)
and the Gaussian width, $\sigma = \sigma_{r^*_{out}} = \sigma_{r^*_{side}} = \sigma_{r^*_{long}}$. 
Both fit parameters from all four correlation functions are in agreement within
statistical errors; combined they
 give $\sigma = 12.5 \pm 0.4  _{-3}^{+2.2} $ fm  and 
$  \langle \Delta r^*_{out}  \rangle= -5.6 \pm 0.6  _{-1.3}^{+1.9}$ fm with a $\chi^{2}$ / dof = 134.5/110.
Systematic errors are estimated from the discrepancy between
the four correlation functions,  the dependence on the input momentum distribution, the uncertainties on 
primary purity and the fit range dependence.
This -5.6 fm in the pair rest frame becomes in the lab frame -3.9 fm (5.4 fm/c) if 
emission difference is space (time) only.



\begin{table}

\begin{tabular}{cccc}

\hline

\hline

\multicolumn{1}{c}{}

&\multicolumn{1}{c}{$\sigma$ (fm)}

&\multicolumn{1}{c}{$\langle \Delta r^*_{out} \rangle$ (fm)}

&\multicolumn{1}{c}{$\chi^{2}$ / dof}\\

\hline

Data & $ 12.5 \pm 0.4  _{-3}^{+2.2} $    & $ -5.6 \pm 0.6  _{-1.3}^{+1.9} $     & 134.5/110 \\

RQMD & $11.8 \pm 0.4$ & $-8.0 \pm 0.6 $ & 205/54\\

\begin{tabular}{c}

RQMD \\ no rescattering \\

\end{tabular}

& $ 5.8 \pm 0.1 $ & $-2.0 \pm 0.3 $ & 940/54 \\ 

BWP & $ 9.9 \pm 0.1$ & $-6.9 \pm 0.3$ & 1020/118 \\ 

\hline\hline

\end{tabular}

\caption{ Fit results using a three dimensional Gaussian distribution in the pair 
rest frame. For the data, the first error is statistical and the second systematic.
The errors on the model calculations are calculated by rescaling the $\chi^2$ distribution by 
the minimum value of $\chi^2$/dof.}

\label{FitComparison}

\end{table}

The parameters $\sigma$ and $\langle \Delta r^*_{out} \rangle$ may be extracted 
directly from BWP or RQMD by constructing
the  $r^* = \sqrt{(r^*_{out})^2 + (r^*_{side})^2 + (r^*_{long})^2}$ and $r^*_{out}$ distributions. However, 
neither RQMD nor BWP $\overrightarrow{r}^*$ 
distribution is close to a three dimensional Gaussian.
Thus, to compare models and data fairly, the correlation functions calculated
from RQMD and BWP are fitted in exactly the
 same way as the data. The extracted fit parameters
are compared to the data in Table 1. The large $\chi^2 / $ dof 
values arise because the tails of the $\overrightarrow{r}^*$ distributions of
RQMD and BWP are not well-described by a three dimensional Gaussian in the pair rest frame.
The data appear to be insensitive to these tails due to larger statistical errors.


Consider BWP. At an emission point, the fluid 
velocity (increases with radius) and the thermal velocity (common freeze-out temperature $T$ for all 
species in fluid rest frame) combine to give the observed particle velocity $\overrightarrow{V}$.  If the 
source does not expand, the relative emission probability for given $\overrightarrow{V}$ will  
track the fireball spatial density. If the source expands but $T=0$, particles with $\overrightarrow{V}$ will 
come from   the single point where the fluid moves with $\overrightarrow{V}$. At $T\ne 0$ and for constant 
density and unlimited fireball size, the spread of thermal velocity smears this emission point to a nearly 
spherical volume whose size increases inversely with particle mass. This volume must be folded with a 
realistic fireball spatial density distribution, removing contributions from large radial distances. 
Thus,  effective centers of emission regions are shifted towards smaller radii. For our $m_t/T$ ($m_t$ = 
transverse mass, and $m_t\propto m$ at given $V$), the relative shift of pions and kaons is small~\cite{LedNonId} 
but measurable. There is also an emission time separation: BWP has kinetic freeze-out at fixed longitudinal 
proper 
time $\tau=\sqrt{t^2-z^2}$, so the larger size of  effective pion source yields emission at later laboratory times 
$t$. Thus pions are on average emitted closer to source center and later in time than kaons. 

In the RQMD model, pion and kaon sources are also spatially shifted when transverse flow builds up by 
hadronic rescattering. Even when rescattering is turned off, resonance decays delay the pion 
average emission time and increase the apparent size of the source.



Our results show that pions and kaons are not emitted
at the same average space-time position for Au+Au collisions at $\sqrt{s_{NN}} = 130$ 
GeV.  The data are consistent with BWP and RQMD, i.e. 
with a system whose dominant feature is a transverse collective expansion.
These results 
significantly challenge models that attempt to explain pion, kaon and proton spectra by purely 
initial state effects~\cite{CGC, InitRescat}.
Such an analysis may also be used to probe at what transverse momentum 
soft processes (expanding system) give way to  hard processes since the
space-time emission pattern will substantially change at that momentum.

We wish to thank the RHIC Operations Group and the RHIC Computing Facility
at Brookhaven National Laboratory, and the National Energy Research 
Scientific Computing Center at Lawrence Berkeley National Laboratory
for their support. This work was supported by the Division of Nuclear 
Physics and the Division of High Energy Physics of the Office of Science of 
the U.S. Department of Energy, the United States National Science Foundation,
the Bundesministerium f\"ur Bildung und Forschung of Germany,
the Institut National de la Physique Nucleaire et de la Physique 
des Particules of France, the United Kingdom Engineering and Physical 
Sciences Research Council, Fundacao de Amparo a Pesquisa do Estado de Sao 
Paulo, Brazil, the Russian Ministry of Science and Technology, the
Ministry of Education of China, the National Natural Science Foundation 
of China, Stichting voor Fundamenteel Onderzoek der Materie, the 
Grant Agency of the Czech Republic, Department of Atomic Energy of India, 
Department of Science and Technology of India, Council of Scientific 
and Industrial Research of the Government of India, and the Swiss National 
Science Foundation.


\end{document}